\documentclass{iau} 

\usepackage{graphicx}
\usepackage{xcolor}
\usepackage[hyphens]{url}
\usepackage{hyperref}
\hypersetup{
     colorlinks  = true,
     linkcolor   = blue, 
     anchorcolor = BrickRed,
     citecolor   = blue,
     filecolor   = blue,
     menucolor   = blue,
     runcolor    = cyan,
     urlcolor    = blue
}
\usepackage[authoryear]{natbib}

\pubyear{2019}
\volume{355} 
\jname{The Realm of the Low-Surface-Brightness Universe}
\editors{D. Valls-Gabaud, I. Trujillo \& S. Okamoto, eds.}

\def\etal{\textit{et al.}}

\title[LSST and the LSB Universe] 
{The Vera Rubin Observatory Legacy Survey of Space and Time \\and the Low Surface Brightness Universe}

\author[S. Brough et al.]   
{Sarah Brough$^{1,2}$, 
Chris Collins$^{3}$, 
Ricardo Demarco$^{4}$, 
Henry C. Ferguson$^{5}$, 
Gaspar Galaz$^{6}$,
Benne Holwerda$^{7}$, 
Cristina Mart{\'\i}nez-Lombilla$^{1,8,9}$, 
Chris Mihos$^{10}$
Mireia Montes$^{1}$
}

\affiliation{$^1$School of Physics, University of New South Wales, NSW 2052, Australia\\
[\affilskip]
$^2$Chair of the LSST Galaxies Science Collaboration Low Surface Brightness Working Group \\ email: {\tt s.brough@unsw.edu.au}\\
$^3$Astrophysics Research Institute, Liverpool John Moores University, Liverpool Science Park, L3 5RF, United Kingdom\\
$^4$Department of Astronomy, Universidad de Concepci\'{o}n, Concepci\'{o}n, Chile\\
$^5$Space Telescope Science Institute, 3700 San Martin Drive, Baltimore, MD 21218, USA\\
$^6$Instituto de Astrof{\'\i}sica, Pontificia Universidad Cat\'{o}lica de Chile, Santiago, Chile \\
$^7$Department of Physics \& Astronomy, University of Louisville, Louisville, KY 40292, USA\\
$^8$Instituto de Astrof{\'\i}sica de Canarias, La Laguna, E-38205, Spain\\
$^9$Departamento de Astrof{\'\i}sica, Universidad de La Laguna, La Laguna, E-38200, Spain\\
$^{10}$Department of Astronomy, Case Western Reserve University, Cleveland, OH 44106, USA\\
}

\begin{document}

\maketitle

\begin{abstract}
The 8.4m Vera Rubin Observatory Legacy Survey of Space and Time (LSST) will start a ten-year survey of the southern hemisphere sky in 2023. LSST will revolutionise low surface brightness astronomy. It will transform our understanding of galaxy evolution, through the study of low surface brightness features around galaxies (faint shells, tidal tails, halos and stellar streams), discovery of low surface brightness galaxies and the first set of statistical measurements of the intracluster light over a significant range of cluster masses and redshifts.
\end{abstract}

\firstsection 

\section{Introduction}

Establishing how galaxies evolve is a crucial component currently missing in our understanding of the Universe. The hierarchical structure formation paradigm requires that every galaxy undergoes interactions throughout its lifetime. However, the precise details of the quantity, frequency or magnitude of those interactions are still not completely understood. We know that each interaction leaves visible signatures around galaxies in faint shells, tidal tails, halos and stellar streams. These low-surface-brightness features around galaxies encode the hierarchical nature of cosmological structure formation (e.g. \citealt{Kaviraj14,Kaviraj19}). However, 
current wide-field imaging surveys like the Sloan Digital Sky Survey (SDSS; \citealt{Abazajian09}) 
become incomplete at $r$-band surface brightnesses fainter than $\mu_{e,r}\sim23.5$ mag arcsec$^{-2}$ \citep{Baldry08}. This means that our current observational understanding of galaxies, and their evolution, is biased to the high-surface brightness galaxies and the merger signatures that are brighter than that (e.g. \citealt{Martin19}).

From the early photographic plate observations of, for example, \cite{Malin80}, it was clear that galaxies show merger-related features at very low surface brightnesses $\mu_{r}>27$ mag arcsec$^{-2}$, and that when surface brightness limits are decreased, we observe many more interesting features around galaxies.  However, this is challenging work. These low surface brightnesses are more than a factor of 100 fainter than the dark night sky (e.g. \citealt{Patat03}) and compete with zodiacal light, galactic cirrus and instrument related issues such as scattered light from the telescope and ghosts. 

Over the last decade CCD imaging has improved to the point that targeted observations have been able to reach these low surface brightnesses. Several different approaches have been taken, from drift scanning \citep{Gonzalez00}, special coatings of the filters \citep{Mihos05}, low surface-brightness-optimized observing techniques with large field-of-view mosaic cameras \citep{Tyson88,Ferrarese12, Duc15, Iodice16} and low surface-brightness-optimized data reduction \citep{Jablonka10,Trujillo16} or even the construction of new low surface-brightness-dedicated telescopes (\citealt{vanDokkum14} and the Huntsman Telescope\footnote{https://tinyurl.com/y3nzlwnd}). 

\cite{Trujillo16} show what is revealed as the surface brightness limit (measured to $3\sigma$ in a $10^{\prime\prime}\times10^{\prime\prime}$ window) is decreased from $\mu_{3,10,10}=26.5$ mag arcsec$^{-2}$ for SDSS, to 
their analysis of Gran Telescopio de Canarias images $\mu_{3,10,10}>31.5$ mag arcsec$^{-2}$, of galaxy UGC 00180. In this example the stellar halo only appears once surface brightness limits reach levels fainter than $\mu_{3,10,10}<29.0$ mag arcsec$^{-2}$ and many surrounding features (like low surface brightness galaxies and tidal streams connecting galaxies) are only visible at these lower surface brightnesses too.

These surveys have also found a significant population of low surface brightness galaxies, including ultra-faint dwarf galaxies \citep{Mihos15b, vanderburg17,Greco18, Mihos18}. This is complementary to the searches for ultra-faint dwarf satellite companions of the Milky Way which have also been found in the Dark Energy Survey \citep{Drlica15} and Hyper Suprime-Cam Subaru Strategic Program \citep{Homma16} through star counts.  There are also extended low surface brightness galaxies with diameters larger than the Milky Way, like Malin 1 \citep{Bothun87}. The volume density of this population of galaxies is still unknown \citep{galaz2015, galaz2019}. Understanding these different populations of galaxies tells us about galaxy formation and the interplay between dark matter and baryonic processes in the cosmological framework of our Universe.

Deep imaging also traces diffuse accretion signatures in the outer disks and halos of nearby galaxies (e.g., \citealt{Mihos13, Watkins15}). The study of the structure of thick discs and truncations in disc galaxies at surface brightness limits of $\mu_r\sim28.5$ mag arcsec$^{-2}$, revealed that a careful point-spread-function (PSF) treatment of observed data is absolutely indispensable when dealing with deep imaging of extended objects \citep{Martinez19}, and that truncations ($R\sim26\pm0.5$ kpc ) are associated with a star formation threshold \citep{Martinez19b}.

This research shows the power of low surface brightness observations for revealing crucial new insights into the Universe and galaxy evolution in particular.  


\subsection{Intracluster Light and Galaxy Evolution}

In particular, our understanding of the evolution of the most massive galaxies in the Universe, Brightest Cluster Galaxies (BCGs), is at an impasse until we can measure the low surface brightness intracluster light (ICL) that they sit within. When observed as an ensemble, the stellar mass of BCGs evolve more slowly at low redshifts $z<0.5$ than predicted by models \citep{Lidman12, Lin13, Oliva14, Zhang15}. This was also seen by \cite{Collins09} but they also found slow growth at higher redshifts that is not observed in more recent analyses. Various models suggest that the slow growth is most likely due to a, currently observationally undefined, fraction of stellar mass joining the ICL in any given merger, rather than joining the final galaxy \citep{Conroy07, Canas19}. The latest generation of galaxy evolution models implement this fraction in interactions
($\sim50$\%) to successfully simulate the observed mass of present-day galaxies and galaxies early in the Universe (e.g. \cite{Guo11, Moster13, Laporte13,Contini18}). It has also been inferred from observations, e.g. \cite{Lidman12, Burke13, Oliva14, Groenewald17}.  
However, 
observations of the ICL to date have only been of small samples ($1-24$ clusters; e.g. \citealt{Gonzalez07, Krick07, Montes14, Burke15, Iodice16, Jimenez16, Mihos17, Morishita17, deMaio18, Montes18}) or employ stacking of many clusters to obtain coarse measurements (e.g. \citealt{Zibetti05, McGee10, Zhang19}). Confusingly, these studies result in very different fractions of intracluster light ($2-80$\%) and contradictory relationships with cluster mass and redshift. Two recent analyses of the ICL in six Frontier Field clusters observed by the Hubble Space Telescope draw very different conclusions on the origins of the ICL, from tidal-stripping of galaxies with stellar masses M$^*\sim6\times10^{10}$ solar masses \citep{Montes18} or M$^*\sim3\times10^{9}$ solar masses \citep{Morishita17}. The cluster mass ranges studied have also been limited: very few systems with masses $10^{13}<$M$_{Cluster}$(solar masses) $<10^{14}$ have been studied to date \citep{daRocha08, deMaio18}.  Therefore, there is no consensus observationally on the fraction of light that joins the ICL over time, or its dependence on cluster mass or redshift.

Unveiling the quantity, origin and growth rate of the ICL, and therefore the efficiency of galaxy interactions in a dense environment, requires a significantly larger sample than has previously been available, over a wide range of cluster mass to quantify the effect the host environment has. See also \cite{Montes19} in these proceedings.

In order to answer all of these questions of galaxy evolution, what do observational astronomers need? We need \emph{DEPTH}, otherwise we will not be able to observe this very faint light. We need a \emph{small, well-measured PSF}, in order to subtract off the galaxies and stars that sit in front and behind the low surface brightness light that we want to measure. We need statistics on these quantities -  a \emph{WIDE} image area is a must. We also need excellent data reduction, to remove all of the sky and telescope signatures that pollute the low surface brightness light.

\section{The Large Synoptic Survey Telescope (LSST)}

The LSST is an 8.4m (6.7m effective) imaging telescope under construction on the Cerro Pachon ridge in Chile. It has 6 filters extending from the ultraviolet to the near-infrared, ($ugrizy$, $320-1050$nm). The LSST camera has a field-of-view of 9.6 deg$^2$. The expected PSF is $\sim0.7^{\prime\prime}$ FWHM ($0.2^{\prime\prime}$/pixel). LSST will undertake a 10 year survey (2023-2033). Of this time, 80-90\% is dedicated to the main Wide-Fast-Deep (WFD) survey and the remaining 10-20\% of time is for mini-surveys and deep drilling fields. 

The motivation behind developing LSST \citep{LSSTScienceBook} was low surface brightness science. The key system performance metric for this is \'{e}tendue -- the product of telescope aperture area in m$^2$ and camera field of view in deg$^2$ as the number of galaxies surveyed to a given surface brightness per unit time is proportional to the \'{e}tendue. LSST has an \'{e}tendue of 319 m$^2$ deg$^2$. The statistical power of a survey of low surface brightness objects is \'{e}tendue $\times$ observing time.

LSST's main Wide-Fast-Deep (WFD) survey will cover $\sim18,000$ deg$^2$ to a 10-year depth of $r\sim27.5$ AB magnitude \citep{Ivezic19}. Over the 10-year nominal survey period, each area on the sky will be visited $50-200$ times in each band. This equates to a potential 10-year surface brightness limit in $r$-band of $>32$ mag arcsec$^{-2}$ ($5\sigma$ in $10^{\prime\prime}\times10^{\prime\prime}$).

At least four deep fields will also be observed; the deep-drilling fields ELAIS S1, XMM-LSS, E-CDFS, COSMOS. Each field will cover approximately 9.6deg$^2$.  The final details of the deep drilling field observing strategy are yet to be determined, including the dithering and total number of visits. However, the observations will likely reach $r=28.5$AB magnitude, equivalent to $33$ mag arcsec$^{-2}$.

\section{The Future of Low Surface Brightness Astronomy with LSST} 

LSST is ideally set-up with the \'{e}tendue and observing time to provide the imaging depth, small PSF and wide area of sky that we need to transform the study of low surface brightness features in galaxy evolution. However, we know that deep imaging observations have intrinsic limitations. These include the contamination of the stellar haloes of galaxies by extended ghost reflections, and the cirrus emission from Galactic dust. Default wide-area imaging pipelines are also optimised to measure the photometry of stars and galaxies (i.e. bright objects extending over small spatial scales), so typically misidentify and over-subtract low surface brightness light. In addition, the ICL is located in highly dense regions and so requires the careful subtraction of the light contributed by the many galaxies present, as well as foreground stars.

Low surface brightness light is more than 100-times fainter than the dark night sky and, in massive low redshift clusters and local galaxy streams, extends over many degrees of sky. Measuring this extended, faint light is technically challenging particularly using default wide-area imaging pipelines. The LSST data pipeline is being actively tested on the Hyper Suprime-Cam Subaru Strategic Program \citep{Aihara19}, including the UltraDeep-COSMOS field which reaches the depth of $i\sim28$ mag at $5\sigma$ for point sources - the expected depth that LSST will reach for the whole sky.

HSC-SSP is no exception in struggling with the known intrinsic limitations \citep{Aihara18} but is also a work in progress with only two data releases having been made to date. For LSST the plan is to retain low surface brightness light by not subtracting the sky background prior to co-adding the multiple observations. Sky background subtraction will need to be exemplary and many groups are working on improved sky background estimators optimized for low surface brightness imaging (e.g. \citealt{Ji18}), as well as improved detection and segmentation techniques (e.g. \citealt{Akhlaghi15, Melchior18, Robotham18}).

Galactic cirrus in particular will remain. Working group members are actively working on this issue - potentially using the difference between the colours of starlight and the galactic cirrus to aid its identification \citep{Roman19}.

Another challenge for LSST will be the identification of low surface brightness features in the large samples as most studies to-date have used visual identification. Working group members are also testing this, including using convolutional neural networks to detect faint tidal features previously visually classified \citep{Atkinson13} in Canada-France-Hawaii Telescope Legacy Survey images \citep{Walmsley19}. Machine-learning offers a promising approach to effective automatic identification of low surface brightness features in and around galaxies. 

First engineering light with LSST's commissioning camera will be in 2021. First light with the full LSST Cam will be in early 2022 so now is the time to test LSST algorithms, and alternative solutions, on low surface brightness data and simulations to check whether the planned approaches work for the particular needs of low surface brightness science. 

\subsection{LSST and the Intracluster Light}

LSST's large field of view will see an average of 35 galaxy clusters at $z<0.1$ per pointing \citep{mak11}. Therefore, over the whole sky LSST will likely find, to $z\sim1.2$: $\sim100,000$ massive clusters and $\sim1$ million groups. This will mean that, for the first time, a facility exists to undertake a statistically significant, quantitative study of the amount, origin and growth rate of intracluster light. This will finally answer the question of the true stellar mass growth rate of BCGs and provide a real test of galaxy evolution simulations.

\section{Where next?}

The LSST Galaxies Science Collaboration, and its constituent Low Surface Brightness Working Group is actively preparing to undertake low surface brightness science with LSST. The roadmap for the Science Collaboration's efforts in the buildup to first light in 2023 is presented in \cite{Robertson19} and \cite{Robertson17}.  We still have more work to do to prepare. Members of LSST are welcome to apply to join our work through the Galaxies Science Collaboration (https://galaxies.science.lsst.org).

There are also many synergies between LSST and other future facilities e.g.: ESA/Euclid whose high-resolution, whole-sky optical photometry and near-infrared photometry and spectroscopy could enable improved cluster detection/mass measurement; NASA/WFIRST whose high-resolution optical and near-infrared observations, ideally of the LSST Deep Drilling Fields, will be able to observe stellar haloes surrounding galaxies and stellar streams out to 40Mpc. As much optical spectroscopy as practical to ensure distances are known as accurately as possible. The recently launched eROSITA X-ray satellite will improve on the depth of ROSAT by a factor of 3 and has a wider energy range than ROSAT (i.e. 0.2-10 keV). The expectation is that it will detect $\sim100,000$ clusters up to $z=1$ and hundreds more up to $z=1.5$. It would also be possible to directly compare LSST low surface brightness measurements with combined TESS imaging for very local sources \citep{Holwerda18}.

\begin{discussion}

\discuss{F. Buitrago}{Given the expected large surface density of objects, will blending prevent separating high-redshift objects from low-redshift ones?}
\discuss{S. Brough}{Blending of targets will be a problem but LSST's Data Management team are working on improved de-blending algorithms (e.g. SCARLET; \citealt{Melchior18}) that are being tested by LSST Science Collaboration members.}

\discuss{D. Valls-Gabaud}{As shown by Trujillo \& Fliri for the GTC,
sampling properly the PSF in the wings
requires rotational offsets and dithers. Can the field be rotated?}
\discuss{S. Brough}{Yes, the LSST field can be rotated but it is not yet clear how rotation will be used in the main survey strategy.}

\discuss{N. Brosch}{Are Constellation-type groups of satellites an issue given their contamination?}
\discuss{S. Brough}{Potentially yes, but both LSST's Chief Scientist and Data Management team are working to mitigate this at different levels.}

\end{discussion}


\begin{thebibliography}{99}

\bibitem[Abazajian \etal(2009)]{Abazajian09}
{Abazajian K.~N., et al.} 2009, \textit{The Astrophysical Journal Supplement Series}, 182, 543

\bibitem[Aihara \etal(2018)]{Aihara18}
{Aihara H., et al.} 2018, \textit{Publications of the Astronomical Society of Japan}, 70, S8

\bibitem[Aihara \etal(2019)]{Aihara19}
{Aihara H., et al.} 2019, \textit{arXiv}, arXiv:1905.12221

\bibitem[Akhlaghi \& Ichikawa(2015)]{Akhlaghi15} 
{Akhlaghi M., Ichikawa T.} 2015, \textit{The Astrophysical Journal Supplement Series}, 220, 1

\bibitem[Atkinson, Abraham \& Ferguson (2013)]{Atkinson13} 
{Atkinson A.~M., Abraham R.~G., Ferguson A.~M.~N.} 2013, \textit{The Astrophysical Journal}, 765, 28

\bibitem[Baldry \etal(2008)]{Baldry08}
{Baldry I.~K., Glazebrook K., Driver S.~P.} 2008, \textit{Monthly Notices of the Royal Astronomical Society}, 388, 945

\bibitem[Bothun \etal(1987)]{Bothun87} 
{Bothun G.~D., Impey C.~D., Malin D.~F., Mould J.~R.} 1987, \textit{The Astronomical Journal}, 94, 23

\bibitem[Burke \& Collins (2013)]{Burke13} 
{Burke C., Collins C.~A.} 2013, \textit{Monthly Notices of the Royal Astronomical Society}, 434, 2856

\bibitem[Burke \etal(2015)]{Burke15}
{Burke C., Hilton M., Collins C.} 2015, \textit{Monthly Notices of the Royal Astronomical Society}, 449, 2353

\bibitem[Ca{\~n}as \etal(2019)]{Canas19}
{Ca{\~n}as R., Lagos C. del P., Elahi P.~J., Power C., Welker C., Dubois Y., Pichon C.} 2019, \textit{arXiv}, arXiv:1908.02945

\bibitem[Collins, \etal(2009)]{Collins09} 
{Collins C. A., et al.} 2009, \textit{Nature}, 458, 603

\bibitem[Conroy, Wechsler \& Kravtsov (2007)]{Conroy07} 
{Conroy C., Wechsler R.~H., Kravtsov A.~V.} 2007,  \textit{The Astrophysical Journal}, 668, 826

\bibitem[Contini, Yi \& Kang (2018)]{Contini18} 
{Contini E., Yi S.~K., Kang X.} 2018, \textit{Monthly Notices of the Royal Astronomical Society}, 479, 932

\bibitem[Da Rocha, Ziegler \& Mendes de Oliveira (2008)]{daRocha08} 
{Da Rocha C., Ziegler B.~L., Mendes de Oliveira C.} 2008, \textit{Monthly Notices of the Royal Astronomical Society}, 388, 1433

\bibitem[deMaio \etal(2018)]{deMaio18}
{DeMaio T., Gonzalez A.~H., Zabludoff A., Zaritsky D., Connor T., Donahue M., Mulchaey J.~S.} 2018, \textit{Monthly Notices of the Royal Astronomical Society}, 474, 3009

\bibitem[Duc \etal(2015)]{Duc15}
{Duc P.-A., et al.} 2015, \textit{Monthly Notices of the Royal Astronomical Society}, 446, 120

\bibitem[Ferrarese \etal(2012)]{Ferrarese12}
{Ferrarese L., et al.} 2012, \textit{The Astrophysical Journal Supplement Series}, 200, 4

\bibitem[Galaz \& Frayer (2019)]{galaz2019} 
{Galaz, G. \& Frayer, D. } 2019, this proceedings

\bibitem[Galaz et al. (2015)]{galaz2015} 
{Galaz, G. et al. } 2015, \textit{The Astrophysical Journal}, 815, L29

\bibitem[Gonzalez \etal(2000)]{Gonzalez00}
{Gonzalez A.~H., Zabludoff A.~I., Zaritsky D., Dalcanton J.~J.} 2000, \textit{The Astrophysical Journal}, 536, 561

\bibitem[Gonzalez \etal(2007)]{Gonzalez07}
{Gonzalez A.~H., Zaritsky D., Zabludoff A.~I.} 2007, \textit{The Astrophysical Journal}, 666, 147

\bibitem[Greco \etal(2018)]{Greco18}
{Greco J.~P., et al.} 2018, \textit{The Astrophysical Journal}, 857, 104

\bibitem[Groenewald, et al. (2017)]{Groenewald17} 
{Groenewald D.~N., Skelton R.~E., Gilbank D.~G., Loubser S.~I.} 2017, \textit{Monthly Notices of the Royal Astronomical Society}, 467, 4101

\bibitem[Guo \etal(2011)]{Guo11} 
{Guo Q., et al.} 2011, \textit{Monthly Notices of the Royal Astronomical Society}, 413, 101

\bibitem[Drlica-Wagner \etal(2015)]{Drlica15} 
{Drlica-Wagner A., et al.} 2015, \textit{The Astrophysical Journal}, 813, 109

\bibitem[Holwerda (2018)]{Holwerda18} 
{Holwerda B.~W.} 2018, \textit{Research Notes of the American Astronomical Society}, 2, 53

\bibitem[Homma \etal(2016)]{Homma16} 
{Homma D., et al.} 2016, \textit{The Astrophysical Journal}, 832, 21

\bibitem[Iodice \etal(2016)]{Iodice16} 
{Iodice E., et al.} 2016, \textit{The Astrophysical Journal}, 820, 42

\bibitem[Ivezi{\'c} \etal(2019)]{Ivezic19} 
{Ivezi{\'c} {\v{Z}}., et al.} 2019, \textit{The Astrophysical Journal}, 873, 111

\bibitem[Jablonka \etal(2010)]{Jablonka10} 
{Jablonka P., Tafelmeyer M., Courbin F., Ferguson A.~M.~N.} 2010, \textit{Astronomy \& Astrophysics}, 513, A78

\bibitem[Ji \etal(2018)]{Ji18} 
{Ji I., Hasan I., Schmidt S.~J., Tyson J.~A.} 2018, \textit{Publications of the Astronomical Society of the Pacific}, 130, 084504

\bibitem[Jim{\'e}nez-Teja \& Dupke (2016)]{Jimenez16} 
{Jim{\'e}nez-Teja Y., Dupke R.} 2016, \textit{The Astrophysical Journal}, 820, 49

\bibitem[Kaviraj (2014)]{Kaviraj14} 
{Kaviraj S.} 2014, \textit{Monthly Notices of the Royal Astronomical Society}, 437, L41

\bibitem[Kaviraj, Martin \& Silk (2019)]{Kaviraj19} 
{Kaviraj S., Martin G., Silk J.} 2019, \textit{Monthly Notices of the Royal Astronomical Society}, 489, L12

\bibitem[Krick \& Bernstein (2007)]{Krick07}
{Krick J.~E., Bernstein R.~A.} 2007,  \textit{The Astronomical Journal}, 134, 466

\bibitem[Laporte \etal(2013)]{Laporte13} 
{Laporte C.~F.~P., White S.~D.~M., Naab T., Gao L.} 2013,  \textit{Monthly Notices of the Royal Astronomical Society}, 435, 901

\bibitem[Lidman \etal(2012)]{Lidman12}
{Lidman C., et al.} 2012, \textit{Monthly Notices of the Royal Astronomical Society}, 427, 550

\bibitem[Lin \etal(2013)]{Lin13}
{Lin Y.-T., Brodwin M., Gonzalez A.~H., Bode P., Eisenhardt P.~R.~M., Stanford S.~A., Vikhlinin A.} 2013, \textit{The Astrophysical Journal}, 771, 61

\bibitem[Mak \etal(2011)]{mak11}
{Mak D.~S.~Y., Pierpaoli E., Osborne S.~J.} 2011, \textit{The Astrophysical Journal}, 736, 116

\bibitem[Malin \& Carter (1980)]{Malin80}
{Malin D.~F., Carter D.} 1980, \textit{Nature}, 285, 643

\bibitem[Martin \etal(2019)]{Martin19} 
{Martin G., et al.} 2019, \textit{Monthly Notices of the Royal Astronomical Society}, 485, 796

\bibitem[Mart{\'\i}nez-Lombilla \& Knapen (2019)]{Martinez19}
{Mart{\'\i}nez-Lombilla C., Knapen J.~H.} 2019, \textit{Astronomy \& Astrophysics}, 629, A12

\bibitem[Mart{\'\i}nez-Lombilla, Trujillo \& Knapen (2019)]{Martinez19b} 
{Mart{\'\i}nez-Lombilla C., Trujillo I., Knapen J.~H.} 2019, \textit{Monthly Notices of the Royal Astronomical Society}, 483, 664

\bibitem[McGee \& Balogh (2010)]{McGee10}
{McGee S.~L., Balogh M.~L.} 2010, \textit{Monthly Notices of the Royal Astronomical Society}, 403, L79

\bibitem[Melchior \etal(2018)]{Melchior18} 
{Melchior P., Moolekamp F., Jerdee M., Armstrong R., Sun A.-L., Bosch J., Lupton R.} 2018, \textit{Astronomy \& Computing}, 24, 129

\bibitem[Mihos \etal(2005)]{Mihos05}
{Mihos J.~C., Harding P., Feldmeier J., Morrison H.} 2005, \textit{The Astrophysical Journal}L, 631, L41

\bibitem[Mihos et al.(2013)]{Mihos13} 
{Mihos, J.~C., Harding, P., Rudick, C.~S., et al.} 2013, \textit{The Astrophysical Journal Letters}, 764, L20

\bibitem[Mihos \etal(2015)]{Mihos15}
{Mihos J.~C., et al.} 2015, \textit{The Astrophysical Journal}L, 809, L21

\bibitem[Mihos et al.(2015)]{Mihos15b} Mihos, J.~C., Durrell, P.~R., Ferrarese, L., et al.\ 2015, \textit{The Astrophysical Journal Letters}, 809, L21

\bibitem[Mihos \etal(2017)]{Mihos17}
{Mihos J.~C., et al.} 2017, \textit{The Astrophysical Journal}, 834, 16

\bibitem[Mihos et al.(2018)]{Mihos18} Mihos, J.~C., Carr, C.~T., Watkins, A.~E., et al.\ 2018, \textit{The Astrophysical Journal Letters}, 863, L7

\bibitem[Montes \& Trujillo (2014)]{Montes14}
{Montes M., Trujillo I.} 2014, \textit{The Astrophysical Journal}, 794, 137

\bibitem[Montes \& Trujillo (2018)]{Montes18}
{Montes M., Trujillo I.} 2018, \textit{Monthly Notices of the Royal Astronomical Society}, 474, 917

\bibitem[Montes (2019)]{Montes19} 
{Montes M. } 2019, \textit{arXiv}, arXiv:1912.01616

\bibitem[Morishita \etal(2017)]{Morishita17}
{Morishita T., Abramson L.~E., Treu T., Schmidt K.~B., Vulcani B., Wang X.} 2017, \textit{The Astrophysical Journal}, 846, 139

\bibitem[Moster, Naab \& White (2013)]{Moster13} 
{Moster B.~P., Naab T., White S.~D.~M.} 2013, \textit{Monthly Notices of the Royal Astronomical Society}, 428, 3121

\bibitem[Oliva \etal(2014)]{Oliva14}
{Oliva-Altamirano P., et al.} 2014, \textit{Monthly Notices of the Royal Astronomical Society}, 440, 762

\bibitem[Patat (2003)]{Patat03}
{Patat F.} 2003, \textit{Astronomy \& Astrophysics}, 400, 1183

\bibitem[Robertson \etal(2017)]{Robertson17}
{Robertson B.~E., et al.} 2017, arXiv, arXiv:1708.01617

\bibitem[Robertson \etal(2019)]{Robertson19}
{Robertson B.~E., et al.} 2019,  \textit{Nature Review Physics}, 1, 450

\bibitem[Robotham \etal(2018)]{Robotham18} 
{Robotham A.~S.~G., Davies L.~J.~M., Driver S.~P., Koushan S., Taranu D.~S., Casura S., Liske J.} 2018, \textit{Monthly Notices of the Royal Astronomical Society}, 476, 3137

\bibitem[Roman \etal(2019)]{Roman19}
{Rom{\'a}n J., Trujillo I., Montes M.} 2019, arXiv, arXiv:1907.00978


\bibitem[LSST Science Collaborations \& LSST Project (2009)]{LSSTScienceBook}
{LSST Science Collaborations and LSST Project} 2009, \textit{arXiv}, arXiv:0912.0201

\bibitem[Trujillo \& Fliri (2016)]{Trujillo16}
{Trujillo I., Fliri J.} 2016, \textit{The Astrophysical Journal}, 816, 98

\bibitem[Tyson \& Seitzer (1988)]{Tyson88} 
{Tyson J.~A., Seitzer P.} 1988, \textit{The Astrophysical Journal}, 335, 552

\bibitem[van der Burg \etal (2017)]{vanderburg17} 
{van der Burg R.~F.~J., et al.} 2017, \textit{Astronomy \& Astrophysics}, 607, A79


\bibitem[van Dokkum \etal(2014)]{vanDokkum14}
{van Dokkum P.~G., Abraham R., Merritt A.} 2014, \textit{The Astrophysical Journal Letters}, 782, L24

\bibitem[Watkins et al.(2015)]{Watkins15} 
Watkins, A.~E., Mihos, J.~C., \& Harding, P.\ 2015, \textit{The Astrophysical Journal Letters}, 800, L3

\bibitem[Walmsley \etal(2019)]{Walmsley19} 
{Walmsley M., Ferguson A.~M.~N., Mann R.~G., Lintott C.~J.} 2019, \textit{Monthly Notices of the Royal Astronomical Society}, 483, 2968


\bibitem[Zhang \etal(2016)]{Zhang15}
{Zhang, Y. et al.} 2016, \textit{The Astrophysical Journal}, 874, 165

\bibitem[Zhang \etal(2019)]{Zhang19}
{Zhang, Y. et al.} 2019, \textit{The Astrophysical Journal}, 874, 165

\bibitem[Zibetti \etal(2005)]{Zibetti05}
{Zibetti S., White S.~D.~M., Schneider D.~P., Brinkmann J.} 2005, \textit{Monthly Notices of the Royal Astronomical Society}, 358, 949

\end{thebibliography}
\end{document}